\documentclass[conference]{IEEEtran}
\IEEEoverridecommandlockouts
\usepackage{amsmath,amsfonts,amssymb}
\usepackage{algorithm}
\usepackage{algpseudocode}
\usepackage{amsmath}
\usepackage{amsthm}
\usepackage{subcaption}

\newtheorem{theorem}{Theorem}

\usepackage{array}
\usepackage[caption=false,font=normalsize,labelfont=sf,textfont=sf]{subfig}
\usepackage[most]{tcolorbox}
\usepackage{xcolor}
\usepackage{textcomp}
\usepackage{stfloats}
\usepackage{url}
\usepackage{verbatim}
\usepackage{graphicx}
\usepackage{caption}   
\usepackage{cite}
\usepackage{tikz}
\usepackage{tikz-3dplot}
\usepackage{booktabs}
\usepackage{bm} 
\usepackage{pgfplots}
\usepackage{pgfplotstable}
\usepackage{balance}
\pgfplotsset{compat=1.18}
\usetikzlibrary{pgfplots.fillbetween} 
\makeatletter
\def\ps@IEEEtitlepagestyle{%
  \def\@oddfoot{\relax}
  \def\@evenfoot{\relax}
}
\makeatother

\begin{document}

\title{Non-Linear Precoding via Dirty Paper Coding for Near-Field Downlink MISO Communications}
\author{\IEEEauthorblockN{Akash Kulkarni and Rajshekhar V Bhat\\}
\IEEEauthorblockA{Indian Institute of Technology Dharwad, Dharwad, India\\
Email: \{ee24dp001, rajshekhar.bhat\}@iitdh.ac.in}}

\maketitle
\begin{abstract}
 
In 6G systems, extremely large-scale antenna arrays operating at terahertz frequencies extend the near-field region to typical user distances from the base station, enabling near-field communication (NFC) with fine spatial resolution through beamfocusing. Existing multiuser NFC systems predominantly employ linear precoding techniques such as zero-forcing (ZF), which suffer from performance degradation due to the high transmit power required to suppress interference. This paper proposes a nonlinear precoding framework based on Dirty Paper Coding (DPC), which pre-cancels known interference to maximise the sum-rate performance. We formulate and solve the corresponding sum-rate maximisation problems, deriving optimal power allocation strategies for both DPC and ZF schemes. Extensive simulations demonstrate that DPC achieves substantial sum-rate gains over ZF across various near-field configurations, with the most pronounced improvements observed for closely spaced users.
\end{abstract}

\section{Introduction}

Wireless systems have advanced rapidly to support higher data rates, massive connectivity, and reliable communication \cite{dang2020should}. Looking toward sixth-generation (6G) networks, technologies such as terahertz (THz) communication and extremely large-scale antenna arrays (ELAAs) are expected to offer significant performance gains. Unlike traditional systems that mainly operate in the far-field region with planar wave assumptions, future 6G deployments will often function in the near-field region, where wavefronts are spherical and classical antenna models become inaccurate \cite{liu2024nearfield_tutorial}. This shift demands new modelling and design approaches.

The near-field–far-field boundary is typically approximated as $R_{\mathrm{FF}} \approx 2D^2/\lambda$, where $D$ is the antenna aperture and $\lambda$ is the wavelength \cite{liu2024nearfield_survey}. When distance $d < R_{\mathrm{FF}}$, spherical wavefronts must be considered; for $d > R_{\mathrm{FF}}$, the plane-wave model applies. Due to the large apertures of ELAAs and the use of high frequencies in 6G, this boundary increases substantially. For instance, a $1$\,m array gives a near-field range of about $20$\,m at $3$\,GHz, $200$\,m at $30$\,GHz, and nearly $2$\,km at $300$\,GHz. Thus, 6G systems using large arrays at THz frequencies will predominantly operate in the near-field, highlighting the need for communication strategies beyond conventional far-field assumptions.

\subsection{Literature Survey}
The unique properties of near-field propagation have motivated extensive research into near-field communication (NFC) systems. Early works established the theoretical foundations by characterising near-field sub-regions \cite{monemi2024study} and clarifying the distinctions between Fraunhofer, radial-focal, and non-radiating distances for phased array antennas. These fundamental studies revealed that near-field beamforming enables finite-depth beam focusing rather than conventional angular beamforming with infinite focus \cite{bjornson2021primer}, opening new possibilities for spatially localised transmission. Building upon these principles, comprehensive surveys have systematically addressed key aspects of NFC, including channel modelling for both spatially-discrete and continuous-aperture arrays, performance analysis in terms of degrees of freedom and power scaling laws, and signal processing techniques such as channel estimation and beamforming design \cite{liu2024nearfield_survey, liu2024nearfield_tutorial, liu2023nearfield_what}.

From a channel modelling perspective, researchers have developed sophisticated models that incorporate electromagnetic field theory. The integration of electromagnetic information theory with wireless system design \cite{zhu2024electromagnetic} has led to physically consistent channel representations that account for spherical wavefronts and spatially varying channel gains. Experimental validation through programmable digital coding metasurfaces has demonstrated the feasibility of dual-channel holographic MIMO communications in near-field scenarios \cite{shao2025dual}, while data-driven approaches using deep convolutional neural networks have shown promise in estimating electromagnetic field distributions \cite{chen2024data}. Despite these advances in understanding near-field propagation, the question of optimal spatial resource utilisation remains open.

A key innovation enabled by near-field propagation is location-based multiple access. Unlike conventional spatial division multiple access (SDMA) that relies solely on angular separation in the far field, location division multiple access (LDMA) exploits additional resolution in the distance domain to distinguish users based on their locations (both angle and distance) \cite{wu2023ldma_conf, wu2023sdma_ldma}. Recent work has further enhanced LDMA by hybridising it with non-orthogonal multiple access (NOMA), applying LDMA to well-resolved users while using NOMA for users with poor spatial resolution \cite{rao2024ldma_noma}. However, existing multiuser near-field schemes predominantly employ linear precoding techniques such as zero-forcing (ZF). While computationally efficient, these linear approaches suffer from performance degradation when users are closely spaced due to residual inter-user interference and spatial correlation, particularly in dense multiuser scenarios where the spatial resolution advantage of near-field beamfocusing is most critical.

\subsection{Contributions}
While NFC has been studied extensively, existing works primarily 
focus on linear precoding techniques. To the best of our knowledge, 
this is the first work to investigate and quantify the performance 
gains of non-linear Dirty Paper Coding in near-field multiuser 
scenarios. The main contributions of this paper are:
\begin{itemize}
    \item We investigate the application of Dirty Paper Coding (DPC) to 
  near-field multiuser MISO systems and demonstrate its significant 
  performance advantages over conventional Zero-Forcing precoding 
  by exploiting optimal interference pre-cancellation.
    \item We formulate and solve the sum-rate maximisation problems under 
  both ZF and DPC schemes, deriving optimal power allocation 
  solutions for each approach in the near-field regime.
    \item Through extensive numerical simulations, we provide several key insights into near-field precoding performance: (i) quantification of the sum-rate gains achieved by DPC over ZF across various near-field configurations, (ii) analysis of how user proximity and spatial correlation affect the performance gap between linear and non-linear precoding, and (iii) geometric interpretation of when and why DPC substantially outperforms ZF in spatially correlated near-field channels.
\end{itemize}

\begin{figure}[t]
    \centering
    \begin{tikzpicture}[scale=1.5]
        
        \draw[thin, ->] (0,0,0) -- (2,0,0) node[anchor=north east]{$z$};
        
        \draw[thin, ->] (0,0,0) -- (0,2,0) node[anchor=north west]{$y$};
        
        \draw[thin, ->] (0,0,0) -- (0,0,2) node[anchor=south]{$x$};
    
        \draw[thin, dashed, ->] (0,0,0) -- (0,0,-2);
        \fill (0,0,0) circle (1.5pt);
        \node[anchor=east] at (0,0,0) {$O$};
    

        \foreach \i in {-1.5,-1,-0.5,0,0.5,1,1.5} {
            \fill[blue] (0,-1,\i) circle (1pt);
        }
        \foreach \i in {-1.5,-1,-0.5,0,0.5,1,1.5} {
            \fill[blue] (0,0,\i) circle (1pt);
        }
        \foreach \i in {-1.5,-1,-0.5,0,0.5,1,1.5} {
            \fill[blue] (0,1,\i) circle (1pt);
        }

        \foreach \i in {-1.5,-1,-0.5,0,0.5,1,1.5} {
            \fill[blue] (0,0.5,\i) circle (1pt);
        }

        \foreach \i in {-1.5,-1,-0.5,0,0.5,1,1.5} {
            \fill[blue] (0,-0.5,\i) circle (1pt);
        }

         \fill[red] ({0.3*(0.2)+1}, {0.5-(0.2)}, -2) circle (1pt);

         \fill[red] ({0.3*(0)+1}, {0.5-(0)}, -3) circle (1pt);

         \fill[red] ({0.4*(0.5)+1}, {0.5-(0)}, 0.5) circle (1pt);

        \node[anchor=west] at ({0.4*(0.5)+1}, {0.5-(0)}, 0.5) {$U_k$};
        
         \node[anchor=east] at (0,0,1) {$Tx$};
         \node[anchor=east] at ({0.3*(0)+1}, {0.5-(0)}, -3) {$U_1$};
         \node[anchor=west] at ({0.3*(0.2)+1}, {0.5-(0.2)}, -2) {$U_2$};
         \draw[thin, ->] (0,0,0) -- ({0.3*(0.2)+1}, {0.5-(0.2)}, -2);
          \node[anchor=east] at (0,0,1) {$Tx$};

         \fill[magenta] (1, 0, 0) circle (1pt);
         \fill[magenta] (1.2,0,0) circle (1pt);
         \node[anchor=east] at (1, -0.2, 0) {$U_3$};
         \node[anchor=west] at (1.2,-0.2,0) {$U_4$};

         \fill[olive] (1.6, -0.3, 0) circle (1pt);
         \fill[olive] (1.6,0.3,0) circle (1pt);
         \node[anchor=west] at (1.6, -0.3, 0) {$U_5$};
         \node[anchor=west] at (1.6,0.3,0) {$U_6$};
         \node[anchor = south] at ({0.3*(0.2)+1)/2}, {(0.5-(0.2))/2}, {-2/2}) {$\mathbf{r}$}; 
    \end{tikzpicture}
      \caption{System model for a generic MISO system with K users.}
    \label{fig:genericMIMO}
\end{figure}
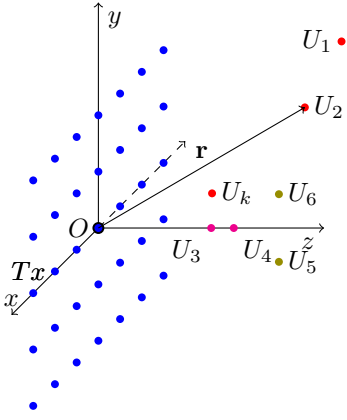

\section{System Model and Problem Formulation}
This section introduces the system model for near-field multiuser MISO downlink transmission, detailing the antenna array configuration, spherical-wave-based channel model, precoded signal model, and the sum-rate maximisation problem under a transmit power constraint.

\subsection{Setup and Near-Field Channel Model}
Consider a base station (BS) equipped with $N$ transmit antennas arranged as a square UPA with $N = N_x \times N_y$, where $N_x = N_y$ represent the number of antennas along the $x$ and $y$ axes, respectively, indexed by $n\in \mathcal{N}\triangleq\{1,2,\ldots,N\}$, serving $K$ single-antenna users in the downlink, indexed by $k\in \mathcal{K}\triangleq \{1,2,\ldots,K\}$, as illustrated in Fig.~\ref{fig:genericMIMO}. Let \( \mathbf{t}^n = [t_x^n,\, t_y^n,\, t_z^n]^\mathrm{T} \in \mathbb{R}^3\) denote the Cartesian coordinates of the \( n \)-th transmit antenna element, and \( \mathbf{r}^k = [r_x^k,\, r_y^k,\, r_z^k]^\mathrm{T} \in \mathbb{R}^3\) denote the coordinates of the \( k \)-th user's receive antenna.

In the line-of-sight (LoS) propagation scenario, the channel coefficient between the $n$-th transmit antenna and the $k$-th user is modelled using the spherical wave propagation as
\begin{equation}\label{eq:channel_coeff}
    h_{k,n} = \frac{1}{\sqrt{4\pi} \| \mathbf{r}^k - \mathbf{t}^n \|} \exp\left(-j \frac{2\pi}{\lambda} \left\| \mathbf{r}^k - \mathbf{t}^n \right\|\right),
\end{equation}
This model accurately captures the spherical wavefront characteristic of near-field propagation, accounting for both amplitude and phase variations across the antenna array. We define the channel matrix $\mathbf{H}\in \mathbb{C}^{K\times N}$, where the $(k,n)$-th element is $h_{k,n}$.

\subsection{Signal Model and Precoding Framework}
Let $\mathbf{s} = [s_1, s_2, \ldots, s_K]^T \in \mathbb{C}^K$ denote the information symbol vector intended for the $K$ users, where $s_k$ represents the symbol for user $k$ with assumption that the symbols  $\mathbb{E}[|s_k|^2] = S$ for all $k \in \mathcal{K}$. 

The transmit signal vector $\mathbf{x} \in \mathbb{C}^N$ is obtained by applying a precoding function $g(\cdot): \mathbb{C}^K \rightarrow \mathbb{C}^N$ to the symbol vector, i.e., $\mathbf{x} = g(\mathbf{s})$.
For linear precoding schemes such as zero-forcing (ZF), the precoding function takes the form $g(\mathbf{s}) = \mathbf{F}\mathbf{s}$, where $\mathbf{F} \in \mathbb{C}^{N \times K}$ is the linear precoding matrix. For non-linear schemes such as dirty paper coding (DPC), the precoding function involves successive encoding with interference pre-subtraction, as will be detailed in a later section. 

The received signal vector $\mathbf{y} = [y_1, \ldots, y_K]^T \in \mathbb{C}^K$ at the $K$ users is given by
\begin{equation}\label{eq:signal_model}
    \mathbf{y} = \mathbf{H}\mathbf{x} + \mathbf{w} = \mathbf{H}g(\mathbf{s}) + \mathbf{w},
\end{equation}
where $\mathbf{w} \sim \mathcal{CN}(\mathbf{0}, \sigma^2\mathbf{I}_K)$ is the additive white Gaussian noise (AWGN) vector with noise power $\sigma^2$ at each user. Without loss of generality, we normalise the noise variance to $\sigma^2 = 1$.

The transmit power is constrained by
\begin{equation}\label{eq:power_constraint}
    \mathbb{E}\left[\|\mathbf{x}\|^2\right] = \mathbb{E}\left[\|g(\mathbf{s})\|^2\right] \leq P_t,
\end{equation}
where $P_t$ is the maximum available transmit power at the BS.

\subsection{Problem Formulation}
Our objective is to maximise the sum-rate across all users:
\begin{equation}\label{eq:sum_rate_problem}
\begin{aligned}
    \max_{g(\cdot)} \quad & R_{\text{sum}} = \sum_{k=1}^{K} R_k \\
    \text{subject to} \quad & \mathbb{E}\left[\|g(\mathbf{s})\|^2\right] \leq P_t,
\end{aligned}
\end{equation}
where $R_k$ denotes the achievable rate for user $k$, which depends on the precoding scheme employed. In the following sections, we investigate both zero-forcing (ZF) and dirty paper coding (DPC) approaches to solve this optimisation problem and compare their performance in near-field scenarios.

\section{Zero-Forcing: Linear, Post-Nulling Interference Cancellation}\label{sec:zf}
We first consider the conventional linear zero-forcing precoding approach as a baseline for performance comparison. 
In this approach, we consider $\mathbf{x} = \mathbf{F}_{\rm ZF}\mathbf{s}$, where $\mathbf{F}_{\rm ZF}\in \mathbb{C}^{N\times K}$ is the linear precoding matrix given by  $\mathbf{F}_{\rm ZF} = \mathbf{H}^H(\mathbf{HH}^H)^{-1}$, which ensures $\mathbf{H} \mathbf{F}_{\rm ZF} = \mathbf{I}_{K}$. Under this precoding, \eqref{eq:signal_model} can be rewritten as 
 \begin{align}\label{eq-input-output_1}
     \mathbf{y} =  \mathbf{H} \mathbf{F}_{\rm ZF} \mathbf{s} + \mathbf{w}   = \mathbf{s} + \mathbf{w}. 
 \end{align}
Thus, the inter-user interference is completely eliminated and the effective channel for the $k^{\rm th}$ user is given by $y_k = s_k + w_k$, yielding a rate of $R_k = \log_2 \left(1 + q_k \right)$, where $q_k \triangleq \mathbb{E}[|s_k|^2]$ for all $k\in \mathcal{K}$. The sum-rate across all users is $R_{\rm sum} = \sum_{k=1}^{K} \log_2 \left(1 + {q_k} \right)$.

The power constraint in \eqref{eq:power_constraint} can be rewritten as 
\begin{align}
   \mathbb{E}\left[ \|\mathbf{x}\|^2\right] &=  \mathbb{E}\left[\mathrm{trace}(\mathbf{x}\mathbf{x}^H)\right]=     \mathbb{E}\left[\mathrm{trace}(\mathbf{F}_{\rm ZF}\mathbf{s} \mathbf{s}^H\mathbf{F}_{\rm ZF}^H)\right]\nonumber  \\
   &= \mathrm{trace}\left(\mathbf{F}_{\rm ZF}\mathbb{E}\left[\mathbf{s} \mathbf{s}^H\right]\mathbf{F}_{\rm ZF}^H\right) =\mathrm{trace}\left(\mathbf{F}_{\rm ZF}\mathbf{S}\mathbf{F}_{\rm ZF}^H\right) \nonumber\\
   &=\sum_{k=1}^{K} q_k \| \mathbf{f}_k \|^2 \leq P_t,  
\end{align}
where $\mathbf{S} = \mathrm{diag}(q_1, q_2, \ldots, q_K)$ is the diagonal power allocation matrix, $\mathbf{f}_k$ denotes the $k$-th column of $\mathbf{F}_{\rm ZF}$, and $\alpha_k  = \| \mathbf{f}_k \|^2$. 

The sum-rate maximisation problem is formulated as 
\begin{align}\label{eq:zf_optimization}
\begin{aligned}
& \underset{q_1, q_2,\ldots,q_K \ge 0}{\text{maximize}} 
& & \sum_{k=1}^{K} \log_2 \left(1 + {q_k}\right),\\
& \text{subject to}
& & \sum_{k=1}^{K} q_k \alpha_k \le P_t.
\end{aligned}
\end{align}
This is a standard convex optimisation problem with a well-known water-filling solution. The optimal power allocation is 
\begin{align}\label{eq:zf_waterfilling}
q_k^\star = \left[\,\frac{1}{\lambda^\star \alpha_k} - 1 \,\right]_+,
\end{align}
where $[\cdot]_+ = \max(0,\cdot)$ and the Lagrange multiplier $\lambda^\star$ is chosen to satisfy  $\sum_{k=1}^{K} \alpha_k q_k^\star = P_t$ with equality.

\section{Non-Linear Dirty Paper Coding: Optimal Pre-Cancellation}
\label{sec:dpc}
To exploit the known interference structure, we employ Dirty Paper Coding (DPC), where the BS sequentially encodes user messages so that interference from later-encoded users is pre-cancelled via Costa precoding \cite{Caire&Shamai_1}. This non-linear approach achieves superior performance compared to linear precoding, particularly in spatially correlated near-field scenarios.

\subsection{Encoding Order Selection and Channel Decomposition}
The performance of DPC critically depends on the encoding order $\pi: \{1,2,\ldots,K\} \rightarrow \{1,2,\ldots,K\}$, which determines the sequence in which users are encoded.  

For a given encoding order $\pi$, we reorder the rows of the channel matrix as $\mathbf{H}_{\pi} = \mathbf{P}_{\pi}\mathbf{H}$, where $\mathbf{P}_{\pi} \in \{0,1\}^{K \times K}$ is the permutation matrix associated with the ordering $\pi$. Specifically, the $k$-th row of $\mathbf{H}_{\pi}$ equals the $\pi(k)$-th row of $\mathbf{H}$, corresponding to user $\pi(k)$.
We then apply the QR decomposition to the Hermitian transpose:
\begin{equation}
    \mathbf{H}_{\pi}^H = \mathbf{Q}^{(\pi)}\mathbf{R}^{(\pi)},
\end{equation}
where $\mathbf{Q}^{(\pi)} \in \mathbb{C}^{N \times K}$ has orthonormal columns ($(\mathbf{Q}^{(\pi)})^H\mathbf{Q}^{(\pi)} = \mathbf{I}_K$) and $\mathbf{R}^{(\pi)} \in \mathbb{C}^{K \times K}$ is an upper triangular matrix. Consequently, we have
\begin{equation}
    \mathbf{H}_{\pi} = (\mathbf{R}^{(\pi)})^H(\mathbf{Q}^{(\pi)})^H,
\end{equation}
where $(\mathbf{R}^{(\pi)})^H$ is a lower triangular matrix with positive diagonal elements $r_{11}^{(\pi)}, r_{22}^{(\pi)}, \ldots, r_{KK}^{(\pi)}$. 

\subsection{DPC Precoding Structure}
For a given encoding order $\pi$, the DPC precoding matrix is set as
\begin{equation}
    \mathbf{F}_{\text{DPC}}^{(\pi)} = \mathbf{Q}^{(\pi)}.
\end{equation}

Substituting this into \eqref{eq:signal_model} and reordering the received signals according to the encoding order $\pi$, we obtain
\begin{align}
    \mathbf{y}_{\pi} &= \mathbf{H}_{\pi}\mathbf{F}_{\text{DPC}}^{(\pi)} \mathbf{s}_{\pi} + \mathbf{w}_{\pi} \nonumber\\
    &= (\mathbf{R}^{(\pi)})^H(\mathbf{Q}^{(\pi)})^H\mathbf{Q}^{(\pi)} \mathbf{s}_{\pi} + \mathbf{w}_{\pi} \nonumber\\
    &= (\mathbf{R}^{(\pi)})^H\mathbf{s}_{\pi} + \mathbf{w}_{\pi},
\end{align}
where $\mathbf{z}_{\pi} = [z_{\pi(1)}, z_{\pi(2)}, \ldots, z_{\pi(K)}]^T$ for $\mathbf{z}\in \{\mathbf{y},\mathbf{s},\mathbf{w}\}$. This transformation converts the original MISO broadcast channel into an equivalent lower triangular channel, where the signal for user $\pi(k)$ under ordering $\pi$ is given by

\begin{align*}
    y_{\pi(k)} &= \sum_{j=1}^{k} r_{kj}^{(\pi)} s_{\pi(j)} + w_{\pi(k)} \nonumber\\
    &= r_{kk}^{(\pi)} s_{\pi(k)} 
    + \underbrace{\sum_{j=1}^{k-1} r_{kj}^{(\pi)} s_{\pi(j)}}_{\text{interference from users } \pi(1), \ldots, \pi(k-1)} 
    + w_{\pi(k)}.
\end{align*}

\subsection{Successive Encoding with Interference Pre-Cancellation}
The key advantage of DPC is that when encoding user $\pi(k) $'s message, the symbols $s_{\pi(1)}, s_{\pi(2)}, \ldots, s_{\pi(k-1)}$ of previously encoded users are known non-causally at the transmitter. The interference term $\sum_{j=1}^{k-1} r_{kj}^{(\pi)} s_{\pi(j)}$ can therefore be pre-subtracted during the encoding process using Costa precoding without increasing the transmit power \cite{Caire&Shamai_1}.

By encoding users sequentially in the order $\pi(1), \pi(2), \ldots, \pi(K)$, each user's message is encoded while treating all previously encoded users' signals as "known dirt" that can be pre-cancelled. After interference pre-cancellation, the effective channel for user $\pi(k)$ under ordering $\pi$ becomes $y_{\pi(k)} = r_{kk}^{(\pi)} s_{\pi(k)} + w_{\pi(k)}$,
yielding an achievable rate of
\begin{align}
   R_{\pi(k)}^{(\pi)} = \log_2\left(1 + (r_{kk}^{(\pi)})^2 q_{\pi(k)}^{(\pi)} \right),
\end{align}
where $q_{\pi(k)}^{(\pi)} = \mathbb{E}[|s_{\pi(k)}|^2]$ is the power allocated to user $\pi(k)$ under encoding order $\pi$. Although the QR decomposition introduces a sign ambiguity, this does not affect the outcome since we consider only the squared diagonal terms of $\mathbf{R}^{(\pi)}$.

\subsection{Sum-Rate Maximisation with DPC}
For a given order $\pi$, the total sum-rate under DPC is
\begin{equation}
    R_{\text{DPC}}^{(\pi)} = \sum_{k=1}^{K} \log_2\left(1 + (r_{kk}^{(\pi)})^2 q_{\pi(k)}^{(\pi)} \right).
\end{equation}

The power constraint can be expressed as
\begin{align}
    \mathbb{E}\left[\|\mathbf{x}\|^2\right] &= \mathbb{E}\left[\|\mathbf{F}_{\text{DPC}}^{(\pi)}\mathbf{s}_{\pi}\|^2\right] = \mathbb{E}\left[\|\mathbf{Q}^{(\pi)}\mathbf{s}_{\pi}\|^2\right] \nonumber\\
    &= \mathbb{E}\left[\mathbf{s}_{\pi}^H(\mathbf{Q}^{(\pi)})^H\mathbf{Q}^{(\pi)}\mathbf{s}_{\pi}\right] = \mathbb{E}\left[\|\mathbf{s}_{\pi}\|^2\right] \nonumber\\
    &= \sum_{k=1}^{K} q_{\pi(k)}^{(\pi)} \leq P_t,
\end{align}
where we used the fact that $\mathbf{Q}^{(\pi)}$ has orthonormal columns, which preserves the norm.

To maximise $R_{DPC}^{(\pi)}$ for a given ordering $\pi$ subject to the power constraint, we formulate the following  problem:
\begin{align}\label{eq:dpc_optimization}
\begin{aligned}
    \max_{q_{\pi(1)}^{(\pi)}, q_{\pi(2)}^{(\pi)}, \ldots, q_{\pi(K)}^{(\pi)} \ge 0} \quad & \sum_{k=1}^{K} \log_2\left(1 + (r_{kk}^{(\pi)})^2 q_{\pi(k)}^{(\pi)}\right) \\
    \text{subject to} \quad & \sum_{k=1}^{K} q_{\pi(k)}^{(\pi)} \leq P_t.
\end{aligned}
\end{align}

The optimal power allocation for ordering $\pi$ is given by
\begin{align}\label{eq:dpc_waterfilling}
(q_{\pi(k)}^{(\pi)})^\star = \left[\,\frac{1}{\lambda^{(\pi)\star}} - \frac{1}{(r_{kk}^{(\pi)})^2} \,\right]_+,
\end{align}
where $\lambda^{(\pi)\star}$ is chosen to satisfy the power constraint $\sum_{k=1}^{K} (q_{\pi(k)}^{(\pi)})^\star = P_t$ with equality.

The optimal encoding order $\pi^\star$ and corresponding precoding matrix $\mathbf{F}_{\text{DPC}}^{(\pi^\star)}$ are obtained by solving \eqref{eq:dpc_optimization} over all $K!$ orderings and selecting one yielding the maximum sum-rate.
\begin{align}\label{eq:optimal_ordering}
    \pi^\star = \arg\max_{\pi} R_{\text{DPC}}^{(\pi)}.
\end{align}

While exhaustive search over all orderings guarantees optimality, it becomes computationally prohibitive for large $K$ due to the factorial growth in complexity. As a low-complexity alternative, we can employ a greedy channel norm-based ordering, where users are ordered in decreasing order of their channel norms $\|\mathbf{h}_k\|^2$. This heuristic approach significantly reduces complexity to $\mathcal{O}(K \log K)$ while providing near-optimal performance in many practical scenarios. To obtain our numerical results, we use an exhaustive search. 

\section{Analytical Characterisation of Precoding Performance}

Before presenting numerical results, we provide analytical insights into the fundamental performance characteristics of ZF and DPC in near-field scenarios. A key contribution of our formulation in Sections~\ref{sec:zf} and~\ref{sec:dpc} is that the entire performance of each precoding scheme can be distilled into specific channel-dependent quantities through our optimisation. For Zero-Forcing, the sum-rate is completely determined by the precoding penalties $\alpha_k = \|\mathbf{f}_k\|^2$, which directly control the feasible power allocation under the constraint $\sum_{k=1}^K \alpha_k q_k \leq P_t$ as seen in \eqref{eq:zf_optimization}. For DPC, the sum-rate is determined by the squared diagonal elements $(r_{kk}^{(\pi)})^2$ of the lower triangular matrix $\mathbf{R}^{(\pi)}$ from QR decomposition, yielding user rates $\log_2(1 + (r_{kk}^{(\pi)})^2 q_{\pi(k)}^{(\pi)})$ as shown in \eqref{eq:dpc_optimization}. This clean separation enables us to characterise scaling behaviour analytically.

A critical question is: \textit{How do these quantities scale with user proximity in the near-field?} To answer this, we analyse and analyse behaviour in the co-linear configuration, where users lie on the same radial line from the BS. The following theorem characterise the asymptotic scaling laws, where we use standard asymptotic notation: $f(\Delta) = \Omega(g(\Delta))$ means $f$ grows at least as fast as $g$, and $f(\Delta) = \Theta(g(\Delta))$ means $f$ and $g$ have the same growth rate.

\begin{theorem}
\label{thm:scaling}
Consider a two-user co-linear near-field MISO system with a uniform planar array (UPA) of $N = N_x \times N_y$ antennas centred at the origin on the $xy$-plane. Two users are positioned along the $z$-axis at $\mathbf{r}_1 = [0, 0, d]^T$ and $\mathbf{r}_2 = [0, 0, d + \Delta]^T$, where $d > 0$ is fixed and $\Delta > 0$ is the inter-user spacing. As $\Delta \to 0$, uniform spherical wave near-field model:

\textbf{(a)} The ZF precoding penalty satisfies $\alpha_k = \|\mathbf{f}_k\|^2 = \Omega(1/\Delta^2)$ for $k \in \{1, 2\}$.

\textbf{(b)} For DPC with encoding order $\pi(1) = 1, \pi(2) = 2$, the diagonal elements satisfy $(r_{11}^{(\pi)})^2 = \Theta(1)$ and $(r_{22}^{(\pi)})^2 = \Theta(\Delta^2)$.
\end{theorem}
\begin{proof}
See Appendix.
\end{proof}

Theorem~\ref{thm:scaling} reveals the fundamental difference between linear and non-linear precoding in spatially correlated near-field channels. As users approach each other ($\Delta \to 0$), the ZF precoding penalty $\alpha_k$ diverges as $1/\Delta^2$, requiring prohibitively large transmit power to maintain non-zero rates. In contrast, DPC's first encoded user maintains constant effective channel gain independent of $\Delta$, while the second user's gain degrades gracefully as $\Delta^2$ without the catastrophic $1/\Delta^2$ power penalty. These fundamental scaling differences explain DPC's substantially better sum-rate performance for closely spaced users, as validated by our numerical results. The asymmetric channel gains in DPC (especially as $\Delta \to 0$) also explain the asymmetric rate regions observed numerically.

\begin{figure*}[t]  
    \centering
    \begin{subfigure}[t]{0.45\textwidth}
        \centering
        \includegraphics[width=\linewidth]{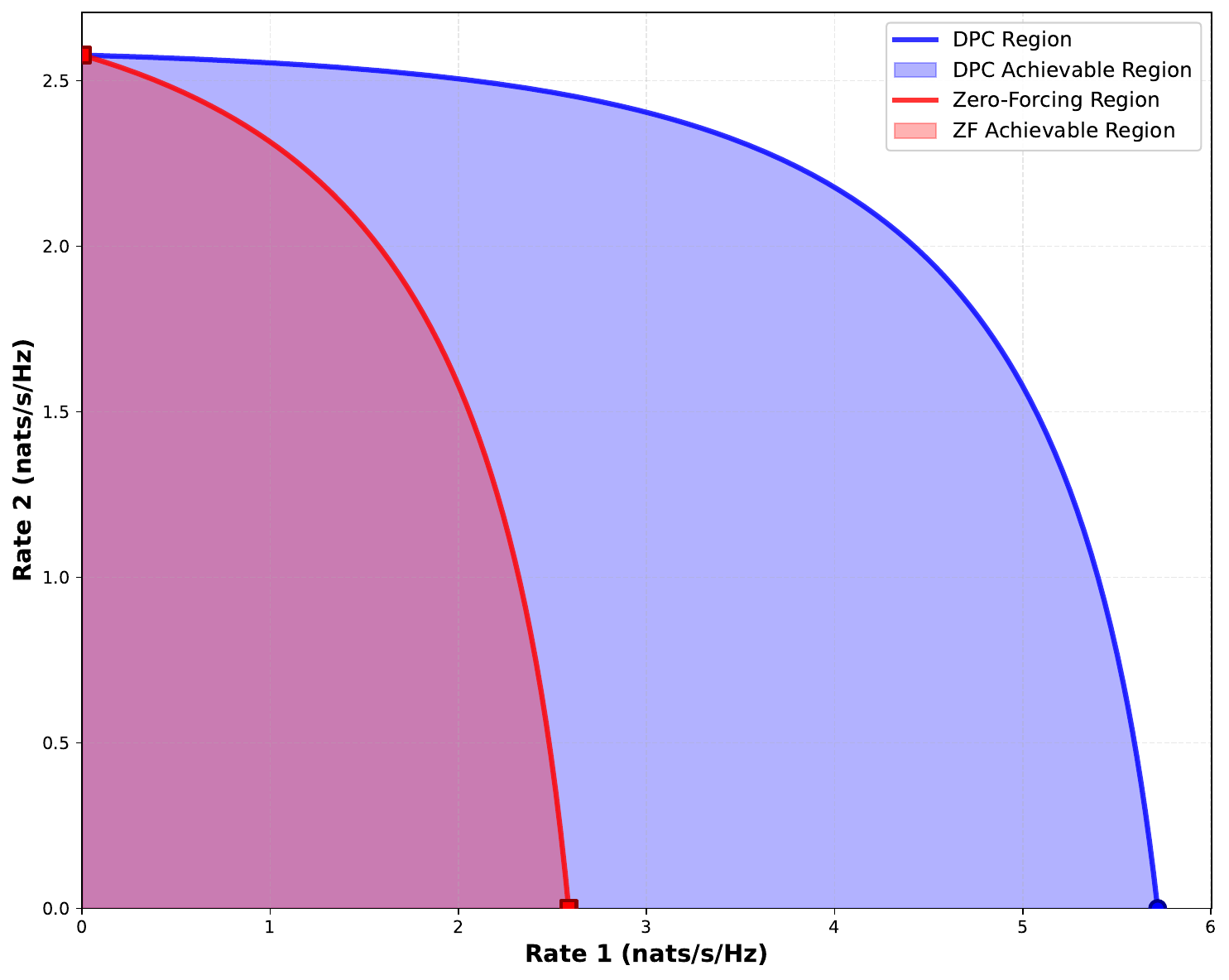}
        \caption{Co-linear: Rate Region: DPC vs ZF}
        \label{fig:Co-linear:Rate Region}
    \end{subfigure}
    \hfill
    \begin{subfigure}[t]{0.45\textwidth}
        \centering
        \includegraphics[width=\linewidth]{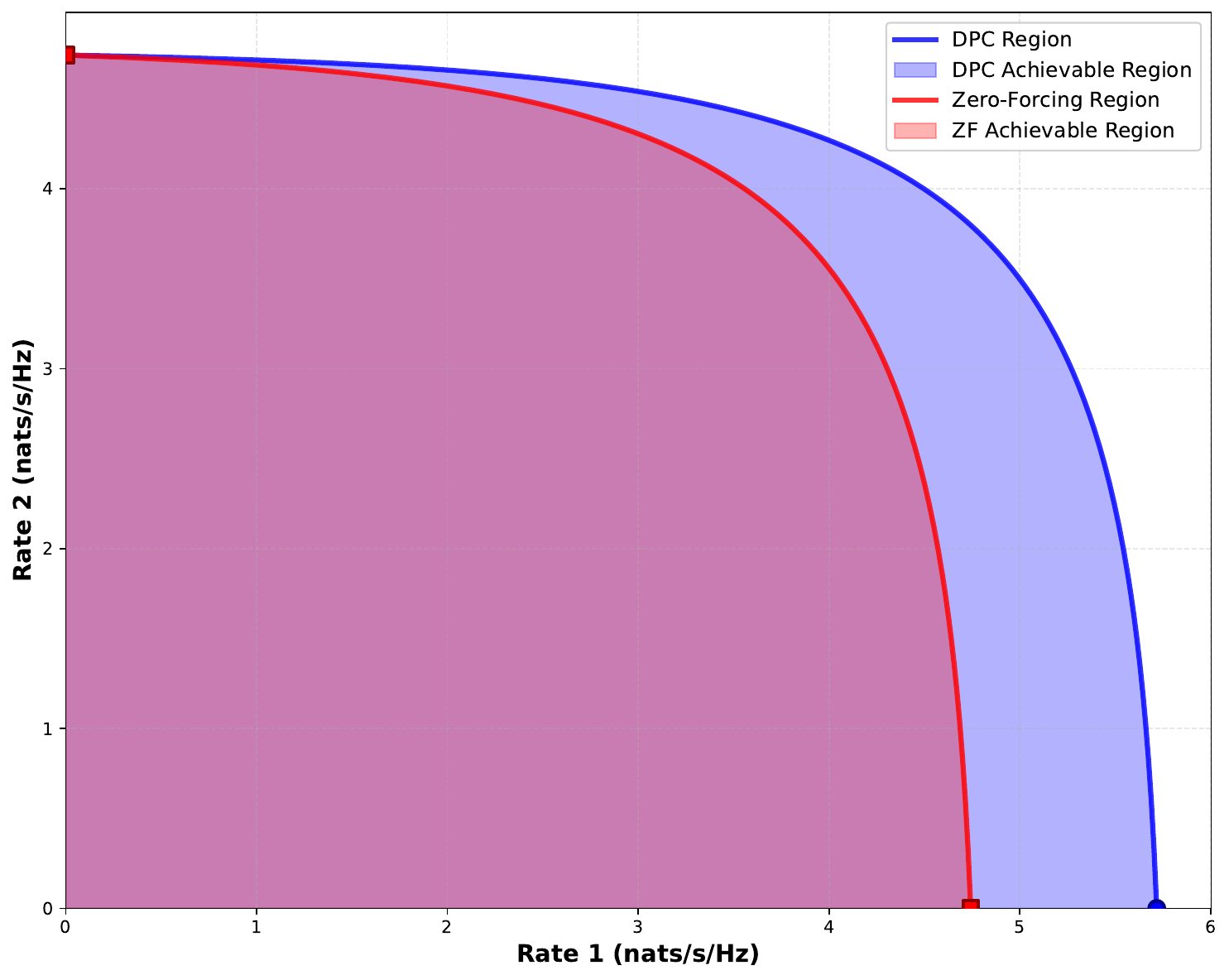}
        \caption{Coplanar: Rate Region: DPC vs ZF}
        \label{fig:Coplaner: Rate Region}
    \end{subfigure}
    \caption{Achievable rate regions comparing DPC and ZF. (a) Co-linear: 
users aligned radially. (b) Coplanar: users separated angularly. 
DPC achieves $147\%$ larger rate region area than ZF in co-linear case.  The parameters chosen are: $N_x= 500$, $d=10$, $P_t =10$, $\Delta =0.2$.  }
    \label{fig:rate-comparison}
\end{figure*}

\section{Numerical Results}

\begin{figure*}[t]  
    \centering
    \begin{subfigure}[t]{0.45\textwidth}
        \centering
        \includegraphics[width=\linewidth]{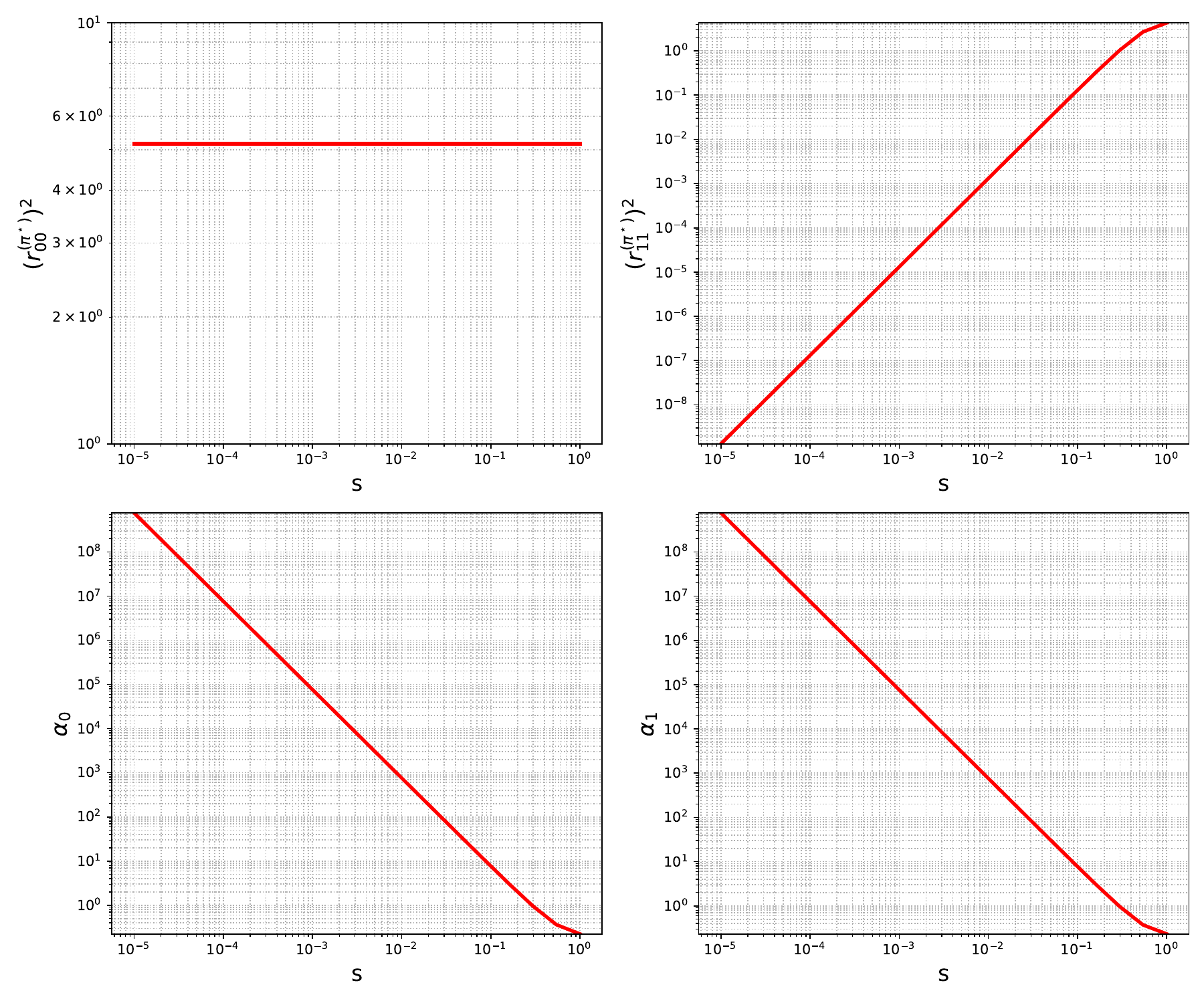}
        \caption{$\alpha_k$ and $(r_{kk}^{(\pi^\star)})^2$}
        \label{fig:plot_output}
    \end{subfigure}
    \hfill
    \begin{subfigure}[t]{0.45\textwidth}
        \centering
        \includegraphics[width=\linewidth]{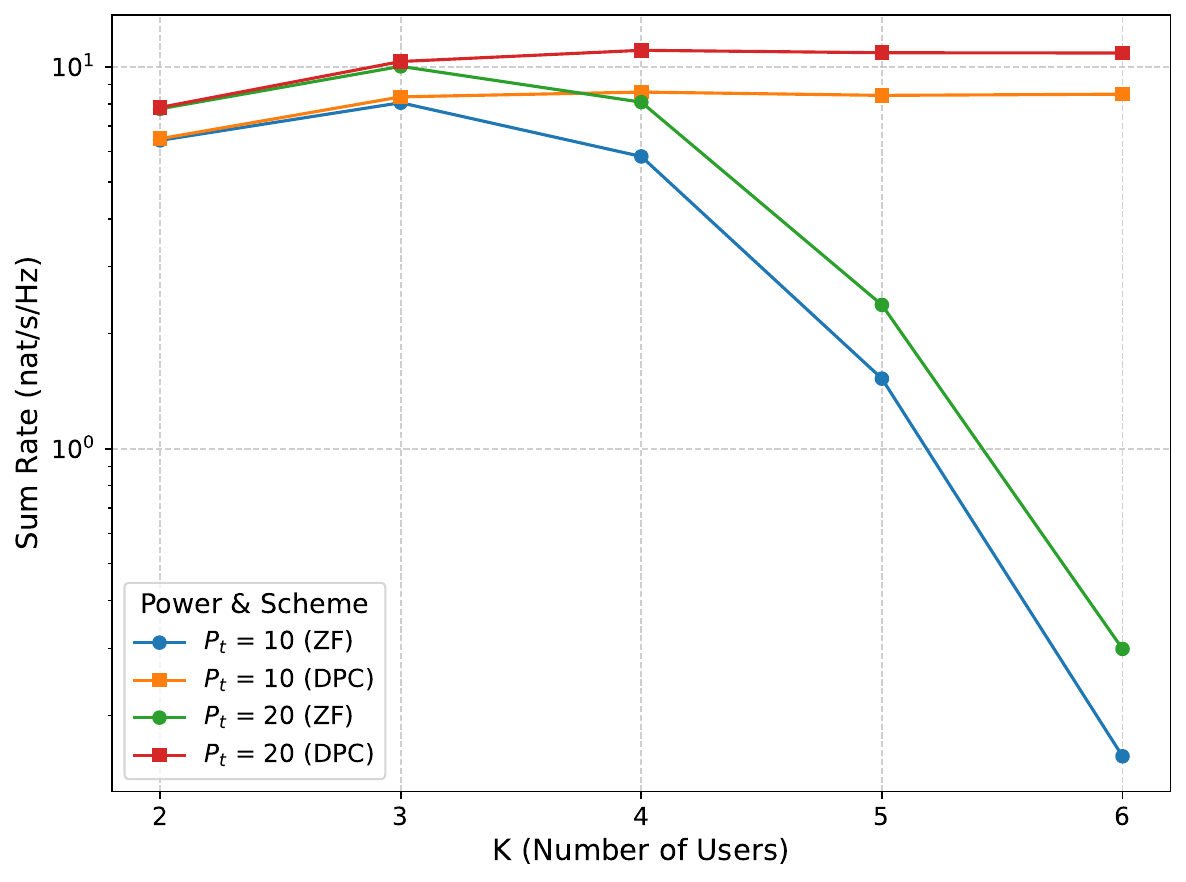}
        \caption{$N_x = 100$}
        \label{fig:SumRate_WS_spacing2}
    \end{subfigure}
    \vspace{0.5cm}
    \begin{subfigure}[t]{0.45\textwidth}
        \centering
        \includegraphics[width=\linewidth]{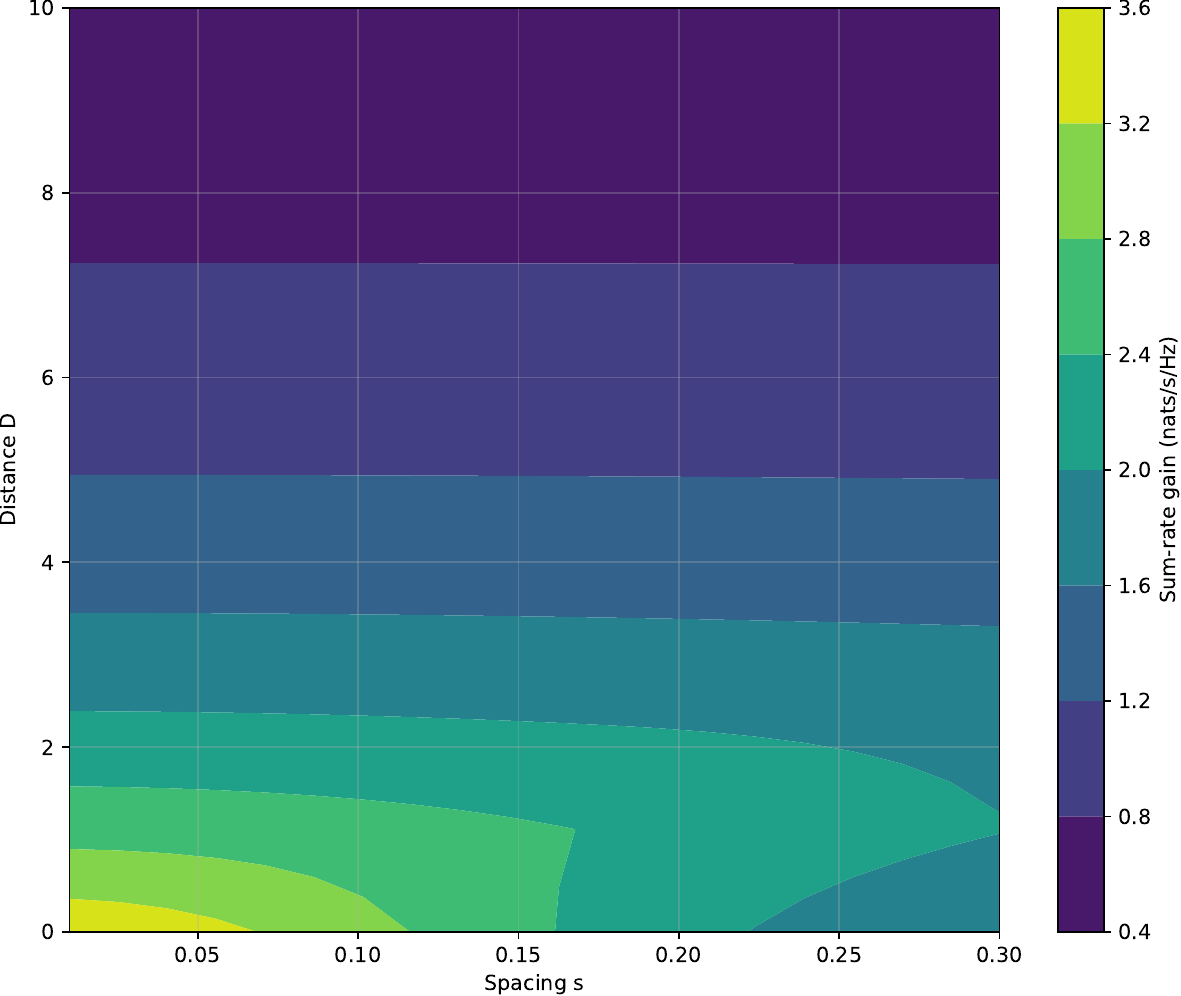}
        \caption{$N_x = 10$}
        \label{fig:contour_data_full_dvss_10_lin_diff}
    \end{subfigure}
    \hfill
    \begin{subfigure}[t]{0.45\textwidth}
        \centering
        \includegraphics[width=\linewidth]{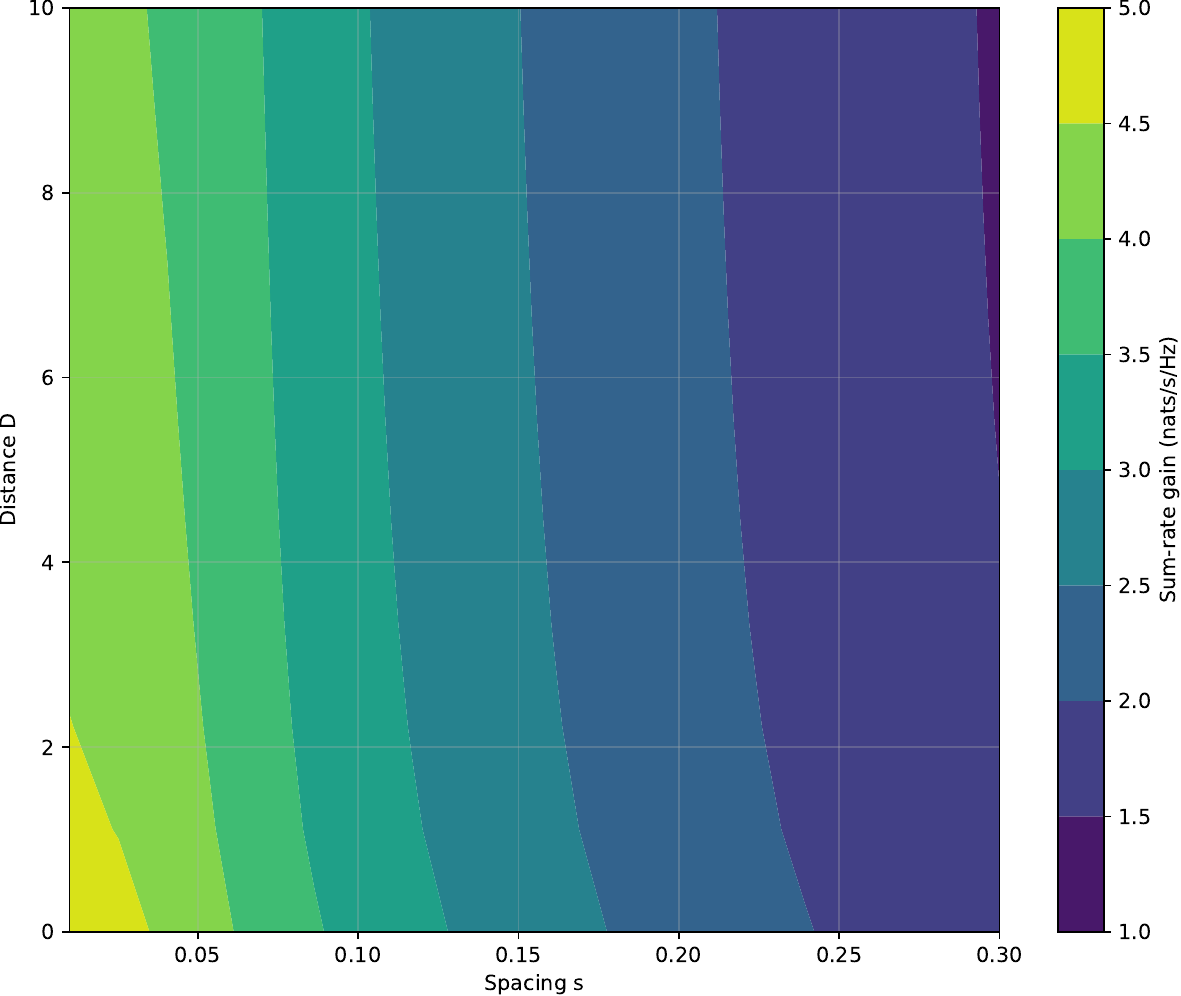}
        \caption{$N_x = 1000$}
        \label{fig:contour_data_full_dvss_1000_lin_diff}
    \end{subfigure}
\caption{(a) Variation of $\alpha_k$ and $(r_{kk}^{(\pi^\star)})^2$ with inter-user spacing $\Delta$. (b) Sum-rate trend for $K$ users arranged colinearly along the $z$-axis within a range of $2$ units. (c), (d) Sum-rate gain comparisons for various distances $d$ and spacings $\Delta$ with $N_x = 10$ and $1000$, respectively.}
    \label{fig:contour-comparison}
\end{figure*}
The wavelength is normalised, and a uniform planar array (UPA) with half-wavelength spacing is centered at the origin on the $xy$-plane. User locations are defined by the distance $d$ from the base station (BS) and the inter-user spacing $\Delta$, either along the line of sight (colinear case) or within the same plane (coplanar case), as shown in Fig.~\ref{fig:genericMIMO}. The colinear setup is more challenging due to user shadowing and high channel correlation, whereas coplanar users offer better angular separation and spatial diversity for effective beamforming.

Fig.~\ref{fig:Co-linear:Rate Region} and Fig.~\ref{fig:Coplaner: Rate Region} show the rate regions for both cases. DPC consistently outperforms ZF by pre-cancelling interference, while ZF benefits more in the coplanar case due to improved channel orthogonality. Notably, DPC yields asymmetric rates in the co-linear case, since successive precoding favours the user encoded first.

Fig.~\ref{fig:plot_output} presents the variation of $\alpha_k$ and $(r_{kk}^{(\pi^\star)})^2$ in a two-user system. As users move closer in the near-field, channel correlation increases, causing $\alpha_k$ to grow sharply under ZF and limiting the available power, which degrades its sum-rate. DPC avoids this power penalty due to its orthonormal precoding structure, but near rank-deficiency in the channel matrix leads to unequal diagonal gains $(r_{kk}^{(\pi^\star)})^2$, giving one user strong performance while the other suffers. Thus, ZF becomes power-inefficient under strong correlation, whereas DPC remains capacity-optimal but may produce highly asymmetric rates when users are nearly co-located. Fig.~\ref{fig:SumRate_WS_spacing2} illustrates the sum-rate trend for $K$ users arranged colinearly along the $z$-axis. The user positions are $10$, $10 + 2/(K-1)$, $10 + 4/(K-1)$, $\ldots$, $12$, assuming a $N_x = 500$ UPA configuration. 
As $K$ increases and users become more closely spaced, DPC maintains or slightly improves the sum-rate by allocating more power to the nearest user, whereas ZF suffers a decline due to increasing difficulty in interference suppression.

Fig.~\ref{fig:contour_data_full_dvss_10_lin_diff} and \ref{fig:contour_data_full_dvss_1000_lin_diff} show the sum-rate gap between DPC and ZF,  with users placed co-linearly over varying $\Delta$ and $d$. The gap is consistently positive, confirming the superiority of DPC. For small arrays, the rate difference is mainly driven by user distance $d$ due to its impact on channel strength. As the array size increases, the influence of $\Delta$ grows, as larger arrays offer finer angular resolution to better separate nearby users. Similar behaviour is observed for coplanar users, where DPC remains more robust to close user spacing by effectively leveraging spatial correlation.

\section{Conclusion}
\label{sec:conclusion}
This paper investigates linear and non-linear precoding schemes for near-field multiuser MISO communications, comparing Zero-Forcing and Dirty Paper Coding with an emphasis on closely spaced user scenarios. Our analysis reveals that DPC consistently outperforms ZF, with rate region area improvements of $147.20\%$ in co-linear and $22.77\%$ in coplanar user configurations. The performance gap between DPC and ZF depends critically on user spacing $\Delta$ and distance $d$ from the BS: the difference decreases as users move farther apart or away from the BS. Array size also plays a role—distance $d$ dominates for small UPAs while spacing $\Delta$ becomes more important for large arrays. DPC's key advantage lies in its interference pre-cancellation, avoiding the power penalty that limits ZF under spatial correlation. As users move closer, ZF suffers from inflated precoder norms and reduced sum-rates, while DPC sustains strong effective channels via successive encoding, albeit with asymmetric user rates. Despite its higher complexity, DPC's robustness in correlated near-field scenarios makes it a strong candidate for future dense deployments.
\balance

\bibliographystyle{IEEEtran}
\bibliography{bibliography}

\appendix
\section*{Proof of Theorem~\ref{thm:scaling}}

We consider a two-user co-linear near-field MISO system with a uniform planar array (UPA) of \(N = N_x N_y\) antennas positioned on the \(xy\)-plane and centered at the origin. The users are located on the \(z\)-axis at \(\mathbf{r}^1 = [0,0,d]^T\) and \(\mathbf{r}^2 = [0,0,d+\Delta]^T\), where \(d>0\) is fixed and \(\Delta>0\) denotes their separation. The \(n\)-th antenna lies at \(\mathbf{t}^n = [t_x^n, t_y^n, 0]^T\), and the wavelength is normalized to \(\lambda=1\).  Throughout the proof we use standard Fresnel-region approximations, together with the uniform spherical wave (USW) channel model. These approximations capture the dependence of the channel vectors on the user spacing~\(\Delta\), which is the quantity of interest in the scaling laws.

The distance from antenna \(n\) to user \(k\in\{1,2\}\) is
\[
d_{kn} = \|\mathbf{r}^k - \mathbf{t}^n\|
       = \sqrt{(t_x^n)^2 + (t_y^n)^2 + (r^k)^2},
\]
where \(r^1 = d\) and \(r^2 = d+\Delta\). Since \(\|\mathbf{t}^n\|\ll d\), the Fresnel expansion yields
\[
d_{kn} \approx r^k + \frac{\|\mathbf{t}^n\|^2}{2r^k}.
\]
Hence,
\[
d_{1n} \approx d + \frac{\|\mathbf{t}^n\|^2}{2d}, 
\qquad
d_{2n} \approx (d+\Delta) + \frac{\|\mathbf{t}^n\|^2}{2(d+\Delta)}.
\]
Under the USW model, the channel coefficients satisfy
\[
h_{1,n} \approx \frac{1}{\sqrt{4\pi d^2}}\,
\exp\!\left(-j2\pi\!\left(d+ \frac{\|\mathbf{t}^n\|^2}{2d}\right)\right),
\]
\[
h_{2,n} \approx \frac{1}{\sqrt{4\pi (d+\Delta)^2}}\,
\exp\!\left(-j2\pi\!\left(d+\Delta+ \frac{\|\mathbf{t}^n\|^2}{2(d+\Delta)}\right)\right).
\]
To relate \(\mathbf{h}_1\) and \(\mathbf{h}_2\), we examine the phase difference
\begin{align*}
\phi_n(\Delta)
&= 2\pi\left[
(d+\Delta)+\frac{\|\mathbf{t}^n\|^2}{2(d+\Delta)}
-\left(d+ \frac{\|\mathbf{t}^n\|^2}{2d}\right)\right] \\
&= 2\pi\!\left[\Delta -\frac{\|\mathbf{t}^n\|^2\Delta}{2d(d+\Delta)}\right].
\end{align*}
For \(\Delta \ll d\),
\[
\phi_n(\Delta) \approx 2\pi\Delta\!\left(1 - \frac{\|\mathbf{t}^n\|^2}{2d^2}\right),
\qquad
\frac{d}{d+\Delta} \approx 1-\frac{\Delta}{d}.
\]
Thus,
\[
h_{2,n} \approx h_{1,n}\, e^{j\phi_n(\Delta)}, 
\qquad
|h_{1,n}|^2 \approx \frac{1}{4\pi d^2}.
\]

The Gram matrix relevant for ZF is
\[
\mathbf{H}\mathbf{H}^H
=
\begin{bmatrix}
 \mathbf{h}_1 \mathbf{h}_1^H & \mathbf{h}_1 \mathbf{h}_2^H \\
 \mathbf{h}_2 \mathbf{h}_1^H & \mathbf{h}_2 \mathbf{h}_2^H
\end{bmatrix},
\]
with
\[
\|\mathbf{h}_1\|^2=\sum_{n=1}^N |h_{1,n}|^2 \approx \frac{N}{4\pi d^2},
\qquad
\|\mathbf{h}_2\|^2 \approx \frac{N}{4\pi d^2}.
\]

Furthermore,
\[
\mathbf{h}_1\mathbf{h}_2^H
\approx \frac{1}{4\pi d^2} \sum_{n=1}^N e^{j\phi_n(\Delta)}.
\]

Expanding
\[
\sum_{n=1}^N e^{j\phi_n(\Delta)}
= e^{j2\pi\Delta}
\sum_{n=1}^N \Bigl[1 - j\frac{\pi\Delta}{d^2}\|\mathbf{t}^n\|^2 + O(\Delta^2)\Bigr],
\]
and letting \(S = \sum_{n=1}^N \|\mathbf{t}^n\|^2\), we obtain
\[
\sum_{n=1}^N e^{j\phi_n(\Delta)}
= N e^{j2\pi\Delta}
\left[1 - j\frac{\pi\Delta S}{N d^2}+O(\Delta^2)\right],
\]
and
\[
|\mathbf{h}_1 \mathbf{h}_2^H|^2
= \left(\frac{N}{4\pi d^2}\right)^2
\left[1+\left(\frac{\pi\Delta S}{N d^2}\right)^2\right] + O(\Delta^3).
\]

\subsection*{(a) Zero-Forcing Case}

The ZF power allocation for user \(k\) is
\[
\alpha_k = \|\mathbf{f}_k\|^2
= \left[(\mathbf{H}\mathbf{H}^H)^{-1}\right]_{kk}.
\]
Hence we require only the diagonal elements of \((\mathbf{H}\mathbf{H}^H)^{-1}\). Substituting the above channel expressions gives
\[
\alpha_k \approx 
\frac{4\pi d^2}{N}
\left(\frac{d^2}{\pi S N}\right)^2
\frac{1}{\Delta^2}
= \frac{4 d^6}{\pi S^2 N^3}\,\frac{1}{\Delta^2}.
\]
Therefore,
\[
\alpha_k = \Omega\!\left(\frac{1}{\Delta^2}\right).
\]
This establishes the result for the ZF case. \qed

\subsection*{(b) DPC Case}

Consider an encoding order \(\pi(1)=1\), \(\pi(2)=2\), and compute the QR decomposition
\[
\mathbf{H}^H = \mathbf{Q}^{(\pi)} \mathbf{R}^{(\pi)},
\]
where \(\mathbf{Q}^{(\pi)}\in\mathbb{C}^{N\times 2}\) has orthonormal columns and \(\mathbf{R}^{(\pi)} \in \mathbb{C}^{2\times 2}\) is upper triangular. The first diagonal entry of \(\mathbf{R}^{(\pi)}\) satisfies
\[
(r_{11}^{(\pi)})^2 
= \|\mathbf{h}_1\|^2 
= \frac{N}{4\pi d^2}
= \Theta(1),
\]
which is independent of \(\Delta\).

The second diagonal entry is
\begin{align*}
(r_{22}^{(\pi)})^2
&= \|\mathbf{h}_2\|^2 
  - \frac{|\mathbf{h}_1^H\mathbf{h}_2|^2}{\|\mathbf{h}_1\|^2} \\
&= \frac{N}{4\pi (d+\Delta)^2}
   - \frac{|\mathbf{h}_1^H\mathbf{h}_2|^2}
          {N/(4\pi d^2)} \\
&= \frac{N}{2\pi d^2}
   + \frac{\pi \Delta^2 S^2}{4 d^4 N}.
\end{align*}
The term depending on \(\Delta\) scales as \(\Delta^2\), and therefore
\[
(r_{22}^{(\pi)})^2 = \Theta(\Delta^2).
\]

This completes the proof of part~(b). \qed

\end{document}